\documentclass[aps,prb,preprint,superscriptaddress]{revtex4-2}
\usepackage{graphicx}  % needed for figures
\usepackage{dcolumn}   % needed for some tables
\usepackage{bm}        % for math
\usepackage{amssymb}
\usepackage[english]{babel}
\usepackage{booktabs}
\usepackage[export]{adjustbox}
\usepackage{array}
\usepackage{subcaption}
\begin{document}
\title{Phonon anharmonicity and soft-phonon mediated structural phase transition in $Cs_3Bi_2Br_9$ }

\author{Debabrata Samanta}
\affiliation{National Centre for High Pressure Studies, Department of Physical Sciences, Indian Institute of Science Education and Research Kolkata, Mohanpur Campus, Mohanpur 741246, Nadia, West Bengal, India.}

\author{Aritra Mazumder}
\affiliation{National Centre for High Pressure Studies, Department of Physical Sciences, Indian Institute of Science Education and Research Kolkata, Mohanpur Campus, Mohanpur 741246, Nadia, West Bengal, India.}

\author{Sonu Pratap Chaudhary}
\affiliation{Department of Chemical Sciences, and Centre for Advanced Functional Materials, Indian Institute of Science Education and Research (IISER) Kolkata, Mohanpur-741246, India.}

\author{Bishnupada Ghosh}
\affiliation{National Centre for High Pressure Studies, Department of Physical Sciences, Indian Institute of Science Education and Research Kolkata, Mohanpur Campus, Mohanpur 741246, Nadia, West Bengal, India.}

\author{Pinku Saha}
\affiliation{National Centre for High Pressure Studies, Department of Physical Sciences, Indian Institute of Science Education and Research Kolkata, Mohanpur Campus, Mohanpur 741246, Nadia, West Bengal, India.}

%\author{Gaurav Shukla}
%\affiliation{Department of Earth Sciences and National Centre for High Pressure Studies, Indian Institute of Science Education and Research Kolkata, Mohanpur Campus, Mohanpur 741246, Nadia, West Bengal, India.}

\author{Sayan Bhattacharyya}
\affiliation{Department of Chemical Sciences, and Centre for Advanced Functional Materials, Indian Institute of Science Education and Research (IISER) Kolkata, Mohanpur-741246, India.}

\author{Goutam Dev Mukherjee}
\email [Corresponding author:]{goutamdev@iiserkol.ac.in}
\affiliation{National Centre for High Pressure Studies, Department of Physical Sciences, Indian Institute of Science Education and Research Kolkata, Mohanpur Campus, Mohanpur 741246, Nadia, West Bengal, India.}
\date{\today}

\begin{abstract} 
We have carried out temperature-dependent x-ray diffraction and Raman scattering experiments on powder $Cs_3Bi_2Br_9$. Trigonal to monoclinic structural transition at around 95 K is discussed and shown to be driven by the softening of the soft mode.  We propose a model to describe the dynamics of the incomplete soft-mode. Raman scattering experiments demonstrate the origin of the soft mode to the rocking motions of Br atoms that participate to form $BiBr_6$ octahedra, which correlates the reported theoretical calculations. Some of the Raman mode frequencies exhibit anomalous temperature dependence due to strong anharmonic phonon-phonon coupling. Temperature-dependent x-ray diffraction analysis estimate the volume thermal expansion coefficient in trigonal phase to be $13.54\times10^{-5} K^{-1}$. In the trigonal phase, the broadening of the full width at half maximum (FWHM) with increase in temperature for $E_g$ and $A_{1g}$ modes is accompanied by  decaying of one optical phonon into two acoustic phonons. The volume thermal expansion rather than anharmonic phonon-phonon interaction dominates the frequency shift for the Raman modes in trigonal phase. In the monoclinic phase, the strength of four phonon processes to the frequency shift and linewidth broadening is much smaller than that for three phonon processes for some of the modes. The observed temperature dependence of FWHM of certain Raman modes in both phases suggests unusual electron-phonon coupling in the crystal.
\end{abstract} 
\maketitle

\section{INTRODUCTION}

During the last few years, halide perovskites of the family $A_3B_2X_9 $   (A is a monovalent cation like Cs or Rb; B is Sb or Bi, and X is a halogen) have gained intense research interest owing to their applications in photovoltaic and optoelectronics devices \cite{Anna,Fan,Bin}.  Due to the importance of these materials, a systematic investigation of their structural, electronic, and optical properties under external stimuli such as pressure and temperature is required.  Some of the $A_3B_2X_9 $ crystals exhibit phase transitions at low temperature as well as at high pressures \cite{IP, Bator, Jakubast, Samanta, Samanta1}. The $Cs_3Bi_2Br_9$ crystal, a member of the aforesaid family has a strong electron-phonon coupling and also undergoes structural transitions at very low pressures \cite{Samanta}.  The structural changes driven by the field environments may hamper device performance. Therefore, it is extremely important to understand the fundamental interactions and correlate structural, electronic, and optical properties of these materials before using them for making devices.

 Temperature-dependent Raman spectroscopic studies are an excellent probe to capture the information about anharmonic interactions of vibrational modes in the crystal. From the variation of Raman linewidth, one can extract information about phonon relaxation processes. The vibrational anharmonicity of the lattice can change the vibrational entropy of the system affecting structural phase transitions, optical properties, etc.  At the onset of a structural phase transition, anomalous changes in full width at half maximum (FWHM),  temperature variation of Raman mode frequency $\omega$ ($d\omega/dT$), and appearance of new Raman modes are usually observed \cite{Miniewicz, Yavalakh}. Ingeneral, Raman mode frequency increases with decrease in temperature.  In certain cases during phase transition Raman mode frequencies show red-shift and go to zero at transition temperatures, which are known as soft modes. The existance of a soft mode suggests structural instability leading to a phase transition \cite{Taniguchi}.
Temperature dependent Raman spectroscopic studies and the variation of complex dielectric permittivity of $Cs_3Bi_2Br_9$ at 1 MHz along a and c directions have revealed a phase transition at 95 K, which is displacive in nature \cite{Bator}. Based on the temperature dependence of nuclear quadrupole resonance (NQR) spectra, heat capacity measurements and low temperature x-ray diffraction (XRD) experiments, Aleksandrova et al. \cite{IP} have reported that the phase transition at 95 K is of second order in nature and is from trigonal ($P\bar{3}m1$) to monoclinic structure (C12/c1). The $Cs_3Bi_2Br_9$ single crystal shows luminescence at 4 K owing to recombinations of free excitons  and excitons bound to defects. The excitation and absorption spectra at that temperature have revealed a direct band gap of energy 2.7 eV \cite{cwm}. A. M Yaremko et al.  \cite{AM} have measured temperature dependence of Raman spectra for few halide perovskites and compared anharmonic interaction of vibrational modes among them. There are very few reports on the temperature evolution of structural and vibrational properties of $Cs_3Bi_2Br_9$. However, the role of thermal expansion, electron-phonon, and phonon-phonon coupling have not been  well investigated yet.

In this work, we have carried out temperature-dependent XRD and Raman spectroscopy measurements to investigate volume thermal expansion, electron-phonon, and anharmonic phonon-phonon interactions. We have also investigated the temperature dependence of  Raman mode frequencies and linewidth using several theoretical model \cite{jaydeep, Ann, Klemens, Balkanski, Hart}, and separated pure temperature and volume thermal expansion contribution to the frequency shift. The brodaning of FWHM with increase in temperature for $E_g$ and $A_{1g}$ modes is accompanied by  decaying of one optical phonon into two acoustic phonons.  We propose a model to describe the dynamic of the soft mode and find that the displacive-type second-order phase transition at 95 K is driven by the softening of the soft mode.

\section{EXPERIMENTAL SECTION}
The $Cs_3Bi_2Br_9$ crystal is prepared by the acid precipitation
method as described in an earlier work \cite{ Samanta}.
Raman spectra are recorded using a confocal micro-Raman spectrometer (Monovista from SI GmBH) in backscattering geometry. A closed cycle cryostat is used for low-temperature Raman spectra measurements. The sample chamber of the cryostat is evacuated using a combination of diaphragm and turbo pump. Temperature is measured using a Cernox sensor placed near the sample. Lakeshore 325 temperature controller is used for temperature control. The micro-Raman spectrometer and cryostat are coupled using 3 plano-convex lenses to collect the scattered light. The glass of the sapphire window inside the cryostat is replaced by two lenses with an anti-reflective coating of a focal length of 50 mm. Another achromatic lens of a focal length of 100 mm is mounted on a linear stage outside the cryostat to adjust the focus. The sample is placed on a copper sample holder attached to the cold head. Raman spectra are collected using a 532 nm laser excitation with laser power $\sim$ 5.4 mW and 2400 $gr/mm$ grating with a resolution of 0.7 $cm^{-1}$. Temperature-dependent XRD measurements are carried out at XRD1 beamline in Elettra synchrotron radiation (Trieste, Italy) with a monochromatic x-ray of wavelength 0.7 $\AA$. The sample is mounted in cryogenic environment and XRD patterns are recorded using a high speed large area detector. All the XRD data are analyzed by GSAS \cite{Brian} program.  

\section{RESULTS AND DISCUSSION}

 We have recorded Raman spectra from 5 $cm^{-1}$ to 250 $cm^{-1}$ in the temperature range from 22 K to room temperature. To investigate the temperature variation of Raman mode frequencies, we have carried out the Lorentz fitting of the modes of the Raman spectrum at each temperature.  At ambient, we have observed eight Raman modes, out of those, six are labeled as $M_1$(29.2 $cm^{-1}$), $M_2$(39.6 $cm^{-1}$), $M_3$(62.9 $cm^{-1}$), $M_4$(65.6 $cm^{-1}$), $M_5$(76.9 $cm^{-1}$), and $M_6$(90.9 $cm^{-1}$) (Fig. 1(a)). The high frequency modes $E_g$ (166 $cm^{-1}$) and  $A_{1g}$(191.1 $cm^{-1}$) originate from the characteristic vibration of Bi-Br bonds in $BiBr_6$ octahedra \cite{AM}. Probably, low-frequency Raman modes are associated to either bending of $Bi-Br$ bonds or rocking motion of $Br$ atoms. The temperature dependence of Raman spectra at selected temperatures is shown in Fig. 1(b).  Certain changes are observed upon decreasing temperature from room temperature down to 22 K. The appearance of a few additional Raman modes below 100 K suggest a change in symmetry of $Cs_3Bi_2Br_9$ crystal.  Increase in number of Raman modes indicate decrease in symmetry, which corraborates the trigonal to monoclinic structural transition as reported by  Aleksandrova et al. \cite{IP}.  Fig. 2. represents Raman spectrum of $Cs_3Bi_2Br_9$ at 22 K. One can see 27 Raman modes at 22 K, which are labeled as $\omega_i$ (where i runs from 1 to 27). The two high-frequency modes at 169 and 194.9 $cm^{-1}$ are probably due to Bi-Br bond stretching and the low-frequency modes are associated with Bi-Br bond bending. We could not carry out the proper assignment of these modes in the absence of polarization measurements, as we have recorded the Raman spectrum on a powder sample.  However, one can see that the intensity of modes $\omega_{i}$ (where i runs from 7 to 15 and 17, 24, 25) are quite low at 22 K. Because of such low intensity, we are unable to carry out the fitting analysis in a reproducible manner. But careful observation tells us, that upon increasing temperature, intensity of all these modes decrease and finally merge with background at higher temperatures. In Fig. 3, we have plotted the temperature variation of selected Raman modes in both phases. In the monoclinic phase all the modes except $\omega_5$ soften with an increase in temperature, whereas the later mode increases with temperature. $\omega_{1}$, $\omega_{3}$, $\omega_{4}$, $\omega_{6}$ and $\omega_{16}$ soften till the monoclinic to trigonal transition temperature approaches. Just below 50 K, $\omega_{2}$ softens and disappears, $\omega_{5}$ and $\omega_{6}$ cross each other, and $\omega_{22}$ merges with $\omega_{21}$ and is red shifted. 
Interestingly, upon decreasing temperature from room temperature both $M_{1}$ and $M_{2}$ modes in the trigonal phase show a red shift up to around 100 K. From Fig. 3(a) it seems that at around 100 K, $M_1$-mode of trigonal phase disappears and reappears as $\omega_4$ mode in monoclinic phase. Therefore, $M_{1}$ mode can be attributed to the characteristic mode of trigonal phase. $M_{2}$ mode keeps on decreasing till the lowest temperature, but with a significant slope change at $T_c$, below which we term it as $\omega_{5}$. The obtained behavior of $M_{1}$ and $M_{2}$ modes may be attributed to strong anharmonic phonon-phonon interaction and are probably directly related to the structural transition.

The frequency of the Raman modes $M_{3}$, $M_{4}$, $M_{5}$, $M_{6}$, $E_{g}$, and  $A_{1g}$ belonging to the trigonal phase are plotted as a function of temperature in Figures 3(b) and 3(c) along with the corresponding close by mode frequency values belonging to the monoclinic structure below around 100 K. All these modes show a blue shift with decreasing temperature with a change in slope across the transition temperature.   $E_g$ and  $A_{1g}$  modes show linear behavior with temperatures above 100 K. Therefore, temperature dependence of these two Raman mode frequencies are analyzed using linear Gru\"{n}eisen model \cite{jaydeep, Ann} 
\begin{equation}
\omega\left(T\right)=\omega_{0}+\chi T,
\end{equation}
\noindent
where $\omega_{0}$ is the Raman mode frequency at 0 K, which is taken as a parameter. $\chi$ is the first order temperature coefficient and the second parameter of the fit. It should be noted that $\chi$ measures the contribution of volume thermal expansion and anharmonic phonon-phonon interactions jointly to the frequency shift. The Eq. (1) is valid only when $2TK_B>>\hbar\omega_0$ \cite {Davydov}. Therefore, the data from 100 K to room temperature are fitted to Eq. (1). Fig. 4(a) shows an excellent fit of the observed data and the fitted parameters are tabulated in Table I. For both the modes, the obtained values of $\chi$ are found to be very close. The distinct slope change in the temperature dependence of frequency for $E_{g}$ and $A_{1g}$ modes below 100 K indicates that the structural transition is possibly initiated by the anharmonic lattice vibration.

Phonon energies can be measured easily but the direct measurement of phonon lifetimes is extremely challenging experimentally \cite{Lory}. However, the temperature dependence of FWHM of a phonon mode can capture a qualitative description of phonon lifetime because FWHM is inversely proportional to the phonon lifetime. The temperature dependence of phonon linewidth gives information about phonon relaxation processes. The variation of phonon linewidth with temperature can be expressed as \cite{Klemens, Balkanski}
\begin{equation}
\Gamma_{j}\left(T\right)=\Gamma_{j}\left(0\right)+A\left[1+\frac{2}{exp\left(\frac{\hbar\omega_{j}\left(0\right)}{2K_{B}T}\right)-1}\right]+B\left[1+\frac{3}{exp\left(\frac{\hbar\omega_{j}\left(0\right)}{3K_{B}T}\right)-1}+\frac{3}{\left(exp\left(\frac{\hbar\omega_{j}\left(0\right)}{3K_{B}T}\right)-1\right)^2}\right],
\end{equation}
where $\Gamma_{j}\left(0\right)$ is the phonon linewidth at 0 K. A and B are anharmonic constants of three and four phonon processes, respectively due to the phonon linewidth broadening. In the trigonal phase, we find that only first two terms of Eq. (2) are sufficient to describe the relaxation prosses for  $A_{1g}$ and $E_g$ modes. Therefore, the FWHM of  $A_{1g}$ and $E_g$ modes are fitted using Eq. (2) with the first two terms and fitted parameters are listed in Table I. Table I shows that the strength of three phonon processes for  $A_{1g}$ mode is much higher than that for $E_g$ mode.
\begin{table}[ht]
	\caption{A list of $\omega_{0}$, $\chi$, $\Gamma_{j}\left(0\right)$, and $A$ for $A_{1g}$ and $E_g$ modes in the trigonal phase.} 
	\centering 
	\begin{tabular}{c@{\hskip 0.2in} c@{\hskip 0.2in} c@{\hskip 0.2in} c@{\hskip 0.2in} c@{\hskip 0.2in}}
		\hline\hline
		Raman mode & $\omega_{j}(0)$ & $\chi$ &$\Gamma_{j}(0)$ & A \\
		& $cm^{-1}$ &  $cm^{-1}/K$ &$cm^{-1}$&$cm^{-1}$\\
		\hline
		$A_{1g}$ & 196.29$\pm$0.05&-0.01837$\pm$0.00032&0.72$\pm$0.06&0.18915$\pm$0.00353 \\
		$E_{g}$ & 170.21$\pm$0.04&-0.01456$\pm$0.00022&0.73$\pm$0.04&0.09509$\pm$0.00223\\
	
		\hline\hline
	\end{tabular}
\end{table} 
The brodaning of FWHM with an increase in temperature for $E_g$ and $A_{1g}$ modes is accompanied by  decaying of one optical phonon into two acoustic phonons. 
%This probably indicates that the structural transition is initiated by the anharmonic lattice vibration.

%{\bf{Where are the values of B in the table? From the fitting it seems that you have to take the B-term for fitting.}}

To describe the strong nonlinear behaviour of low-frequency phonon modes with temperature we need a much better description of  $\omega(T)$. The variation of frequency of phonon mode(j) with temperature can be expressed as
\begin{equation}
\omega_{j}\left(T\right)=\omega_{j}\left(0\right)+\left(\Delta\omega_{j}\right)_{qh}\left(T\right)+\left(\Delta\omega_{j}\right)_{anh}\left(T\right),
\end{equation}
where $\omega_{j}\left(0\right)$ is the frequency of $j^{th}$ phonon mode at 0 K. $\left(\Delta\omega_{j}\right)_{qh}\left(T\right)$  corresponds to the frequency shift from quasiharmonicity and $\left(\Delta\omega_{j}\right)_{anh}\left(T\right)$ corresponds to the frequency shift caused by an anharmonic phonon-phonon coupling. The frequency shift due to an anharmonic phonon-phonon coupling can be modeled as \cite{Balkanski}
\begin{equation}
\left(\Delta\omega_{j}\right)_{anh}\left(T\right)=C\left[1+\frac{2}{exp\left(\frac{\hbar\omega_{j}\left(0\right)}{2K_{B}T}\right)-1}\right]+D\left[1+\frac{3}{exp\left(\frac{\hbar\omega_{j}\left(0\right)}{3K_{B}T}\right)-1}+\frac{3}{\left(exp\left(\frac{\hbar\omega_{j}\left(0\right)}{3K_{B}T}\right)-1\right)^2}\right],
\end{equation}
where the first term in Eq. (4) corresponds to cubic anharmonicity that describes the decay of one optical phonon into two acoustic phonons of equal frequency . The second term describes the decay of one optical phonon into three identical acoustic phonons. C and D are constants, measure the strength of three and four phonon processes to the frequency shift, respectively. The phonon frequency shift caused by quasiharmonicity or thermal expansion contribution can be expressed as \cite{Zhouying, Menendez}
\begin{equation}
\left(\Delta\omega_{j}\right)_{qh}\left(T\right)=\omega_{j}\left(0\right)\left[exp\left(-\int_{0}^{T}\gamma_{j}\left({T}'\right)\beta\left({T}'\right){dT}'\right)-1\right],
\end{equation}
where $\gamma_{j}\left({T}'\right)$ is the mode Gru\"{n}eisen parameter of $j^{th}$ mode and $\beta\left({T}'\right)$ is the volume thermal expansion coefficient. We shall discuss about the temperature dependence of Raman mode frequencies for the moloclinic phase later.\\

To estimate the thermal expansion coefficient we have performed XRD experiments over the temperature range 100-280 K. A few XRD patterns at selected temperatures are presented in Fig. 5. During increasing temperature from 100 K, all XRD peaks shift to lower $2\theta$ due to increasing lattice parameters. Rietveld refinement analysis is employed to calculate cell volume. The atomic coordinates reported by F. Lazarini et al. \cite{Lazarini} are used as the structure model to perform Rietveld refinement of the XRD patterns. The Rietveld refinement plot at 100 K is shown in Fig. 6(a), which shows a good fit with lattice parameters: a=7.58044(9)$\AA$, c=9.4003(3)$\AA$, V=467.81(2)$\AA^{3}$. Fig. 6(b) represents a plot of ln($V$) $\it vs.$ $T$. The volume thermal expansion coefficient is defined as $\beta\left(T\right)=\frac{d(lnV)}{dT}$, where $V$ is the cell volume. We have estimated $\beta\left(T\right)=(13.38\pm 0.39)\times10^{-5} $ above 95 K. It should be mentioned that the thermal expansion coefficient of $Cs_{3}Bi_{2}Br_{9}$ is slightly smaller than that for $Cs_3Bi_2I_9$ \cite{Kyle} ($\beta=15.1\times10^{-5}K^{-1}$), however much larger than copper indium gallium selenide (CIGS) \cite{Jesper} ($\beta=2.7\times10^{-5}K^{-1}$), CdTe  ($\beta=1.4\times10^{-5}K^{-1}$) thin film solar cell materials. The obtained high value of thermal expansion coefficient of $Cs_3Bi_2Br_9$ implies that it will expand more when being heated. Therefore, the effect of thermal expansion to its electronic properties needs to be taken into account when designing optoelectronic devices.\\

 Temperature variation of Raman mode frequencies of $M_{3}$, $M_{4}$, $M_{5}$, and $M_{6}$ in the trigonal phase are analyzed using Eq. (3) without four phonon processes and Fig. 7(a) shows that excellent fit to our data. The fitted parameters are tabulated in Table II. 
% \begin{table}[ht]
%	\caption{A list of $\omega_{j}(0)$, $\gamma$, and C for few Raman modes.} 
%	\centering 
%	\begin{tabular}{c@{\hskip 0.5in} c@{\hskip 0.5in} c@{\hskip 0.5in} c@{\hskip 0.5in} }
%		\hline\hline
%		Raman mode & $\omega_{j}(0)$& $\gamma$&C \\
%		& $cm^{-1}$ &  &$cm^{-1}$\\
%		\hline
%		$M_{3}$ & 66.25$\pm$ 0.06& 1.42 &0.00190$\pm$ 0.00120\\
%		$M_{4}$ & 68.06$\pm$ 0.03& 1.50 &0.01639$\pm$ 0.00072\\
%		$M_{5}$ & 77.46$\pm$ 0.04& 0.97 &0.03343$\pm$ 0.00090\\
%		$M_{6}$ & 93.81$\pm$ 0.04& 0.59 &0.01032$\pm$ 0.00105\\
%		\hline\hline
%	\end{tabular}
%\end{table} 
It is assumed that mode Gru\"{n}eisen parameter is independent of temperature and the values of mode Gru\"{n}eisen parameters are taken from the high pressure studies of $Cs_{3}Bi_{2}Br_{9}$ \cite{Samanta}. We have estimated the individual contribution of volume thermal expansion and anharmonic phonon-phonon coupling to the frequency shift. Fig. 7(b) represents temperature dependence of frequency shift of $M_{3}$, $M_{4}$, $M_{5}$, and $M_{6}$ modes due to the volume thermal expansion and anharmonic phonon-phonon  coupling separately. As expected volume thermal expansion leads to a negative frequency shift for all the Raman modes. Interestingly, anharmonic phonon-phonon coupling lead to the positive frequency shift for $M_{3}$, $M_{4}$, and $M_{5}$ modes and negative frequency shift for $M_{6}$ Raman mode. The low-frequency modes are probably related to the deformation of $BiBr_6$ octahedra and hence show a positive anharmonic contribution to the frequency variation with temperature. It is clearly seen that the frequency shift due to the volume thermal expansion is greater than that for anharmonic phonon-phonon coupling. 
%{\bf{From Fig M5 fit does not look good. Check the 300 K data!}}
To know the information about phonon relaxation processes the temperature dependence of phonon linewidth of $M_{3}$, $M_{4}$, $M_{5}$, and $M_{6}$ modes in trigonal phase are fitted to Eq. (2) without taking into account the third term. (Fig. 7(c))A good fit is obtained for all modes. All the fitting results are listed in Table II.  

%{\bf{It is better to fit all using all terms in Eq.2 and then compare the values}}

  \begin{table}[h]
 	\caption*{Table II: A list of $\omega_{j}(0)$, $\gamma$, C, D, $\Gamma_{j}\left(0\right)$, A, B for few Raman modes.} 
 	\centering 
 	\begin{tabular}{c@{\hskip 0.2in} c@{\hskip 0.2in} c@{\hskip 0.2in} c@{\hskip 0.2in} c@{\hskip 0.2in} c@{\hskip 0.2in} }
 		\hline\hline
 		Raman mode & $\omega_{j}(0)$& $\gamma$&C&$\Gamma_{j}(0)$ & A \\
 		& $cm^{-1}$ &  &$cm^{-1}$& $cm^{-1}$&$cm^{-1}$\\
 		\hline
 		$M_{3}$ & 66.25$\pm$ 0.06& 1.42 &0.00190$\pm$ 0.0012&2.50$\pm$ 0.10&0.07911$\pm$ 0.00211\\
 		$M_{4}$ & 68.06$\pm$ 0.03& 1.50 &0.01639$\pm$ 0.00072&1.60$\pm$ 0.07&0.04377$\pm$ 0.00162\\
 		$M_{5}$ & 77.46$\pm$ 0.04& 0.97 &0.03343$\pm$ 0.00090&1.55$\pm$ 0.04&0.05353$\pm$ 0.00110\\
 		$M_{6}$ & 93.81$\pm$ 0.04& 0.59 &0.01032$\pm$ 0.00105&1$\times 10^{-14}$&0.11894$\pm$ 0.00212 \\
 		\hline\hline
 	\end{tabular}
 \end{table} 
 
Now we discuss phonon anharmonicity in the monoclinic phase. It has been found that the linear thermal expansion coefficient in the monoclinic phase is much smaller than that in the trigonal phase \cite{Bator}.  Therefore, the variation of Raman mode frequencies of $\omega_{18}$ and $\omega_{19}$ with temperature below 100 K (in the monoclinic phase) are fitted to Eq. (3) ignoring the contribution of volume thermal expansion to the frequency shift (Fig. 8). The fitting results show the strength of three phonon processes to the frequency shift is much higher than that for four phonon processes (Table III). (Fig. 8) Temperature dependence of linewidth broadening of these modes are fitted to Eq. (2). In the process of reduction of phonon lifetime of these two modes, there is a much larger probability of occurrence of one optical phonon decaying into two acoustic phonons than into three acoustic phonons (Table III).
 
  \begin{table}[h]
 	\caption{A list of $\omega_{j}(0)$, C, D, $\Gamma_{j}\left(0\right)$, A, B for $\omega_{18}$ and $\omega_{19}$ Raman modes.} 
 	\centering 
 	\begin{tabular}{c@{\hskip 0.4in} c@{\hskip 0.4in} c@{\hskip 0.4in} c@{\hskip 0.4in} c@{\hskip 0.4in} }
 		\hline\hline
 		Raman mode & $\omega_{j}(0)$& C&D\\
 		& $cm^{-1}$ & $cm^{-1}$ &$cm^{-1}$\\
 		\hline
        $\omega_{18}$& 66.01$\pm$0.04 & 0.01061$\pm$0.00560 &-0.00099$\pm$0.00010  \\
         $\omega_{19}$& 68.08$\pm$0.03 & -0.02845$\pm$0.00478 &-0.00013$\pm$0.00009 \\
         \hline\hline
        Raman mode & $\Gamma_{j}(0)$ & A & B\\
        & $cm^{-1}$ & $cm^{-1}$ &$cm^{-1}$\\
        \hline
        $\omega_{18}$& 0.48$\pm$0.09 & 0.21888$\pm$0.01323 &-0.00148$\pm$0.00024  \\
        $\omega_{19}$& 1.92$\pm$0.06 & -0.01027$\pm$0.00898 &0.00094$\pm$0.00017  \\
 		\hline\hline
 	\end{tabular}
 \end{table} 
 
  The $M_1$ mode softens until the transition temperature is reached followed by a hardening down to 22 K, the lowest temperature of the experiment. The behavior of this mode is not exactly similar to the dynamics of a soft mode. The soft mode softens with the decrease in temperature due to the renormalization of the phonon frequency, which is caused by an anharmonic interaction in the crystal. Ideally, a soft mode softens to zero frequency at $T_c$. In the present case, the structural phase transition hinders the softening of the soft mode to zero frequency because the phase transition has occurred at a much higher temperature than that required for softening of the soft mode to zero frequency. Bass et al. \cite{Bass} reported the existence of two soft modes of symmetry $E_g$ and $A_{1u}$ through phonon-dispersion calculation using density functional theory. According to our investigations, the observed soft mode at 29.2 $cm^{-1}$ ($M_1$) is assigned to $E_g$ symmetry. The soft mode originates due to the rocking motions of Br atoms that take part to form $BiBr_6$ octahedra \cite{Bass}. The $A_{1u}$ mode is neither Raman nor infrared active. Therefore, it is expected not to be observed in the Raman scattering experiment. The soft-mode, which does not go to zero frequency at $T_c$, we term it as an incomplete soft-mode. To understand the dynamic of the incomplete soft-mode, we propose a model. We  propose that the onset of structural transition possibly causes appearance of new phonon modes, which may scatter the soft mode. This may results in the observation of the incomplete soft mode. Let us consider the response of a damped harmonic oscillator subjected to a driving force $\frac{F}{m}\cos(\omega t)$:
 \begin{equation}
 	\ddot{x}+\frac{\alpha}{m}\dot{x}+\omega_0^2x=\frac{F}{m}\cos(\omega t),
 \end{equation}
 where $\alpha$ is a damping constant. The natural frequency of the harmonic oscillator is denoted by $\omega_0^2=\frac{k}{m}$. We focus on the situation at large values of t and the condition for maximum energy. A simple calculation based on the aforementioned conditions we obtain 
  \begin{equation}
 	\omega^2=\omega_0^2-\omega_0\Delta,
 \end{equation}
 where $\Delta=\frac{\alpha^2}{2m\sqrt{km}}$ can be correlated to the anharmonicity in the crystal. For the three-phonon process in which one optical phonon decaying into two acoustic phonons with energy $\hbar \omega_1+\hbar\omega_2=\hbar\omega$ and momentum $q_1+q_2=q$, the relaxation rate depends on the factor \cite{Klemens}
   \begin{equation}
 	\Delta=-A(n_1+n_2+1),
   \end{equation}
where A is a constant. $n_1$ and $n_2$ are the occupation numbers at equilibrium. In the approximation when $K_BT>>\hbar\omega$ and $\omega_1=\omega_2=\frac{\omega_0}{2}$, the relaxation rate can be written as
   \begin{equation}
	\Delta  \approx-(\omega_0^{-1}CT+A).
   \end{equation}
 Now consider a double well potential $V(x)=-kx^2+jx^4$ with $k$ and $j$ positive. In the harmonic approximation $\omega_0^2$ would be a negative quantity and we set \cite{Scott}
    \begin{equation}
 	\omega_0^2=-CT_c,
    \end{equation} 
where $C$ and $T_c$ are positive constants. Substituting Eq. (9) and Eq. (10) in Eq. (7) we obtain 
    \begin{equation}
	\omega^2(T)=C(T-T_c) +\omega_c^2.
    \end{equation} 
Eq. (11) describes a soft phonon whose frequency is $\omega_c$ at the transition temperature. We thus obtain an equation describing the dynamics of an incomplete soft mode. Fig. 9 shows an evolution of a soft phonon with temperature. The observed behaviour of the soft phonon is fitted to Eq. (11). The fit gives $\it C$= 0.38(2) $cm^{-2}/K$, $T_c$=100.4(1) K,  $\omega_c^2$=779.7(1) $cm^{-2}$. The estimated transition temperature is in good agreement with the previous literature \cite{IP}. Therefore, we believe the displacive-type second-order phase transition is driven by the softening of the soft mode.

Fig. 10 represents linewidth of $\omega_{1}$, $\omega_{3}$, $\omega_{4}$,   $\omega_{5}$, $M_1$,  and $M_2$ modes as a function of temperature.  A peak-like behavior is seen at 100 K for $\omega_{4}$ and $\omega_{5}$ modes whereas FWHM of  $\omega_{1}$ and $\omega_{3}$ modes increase with increasing temperature followed by a sharp drop at 60 K. The observed behavior of  $\omega_{1}$, $\omega_{3}$, $\omega_{4}$, $\omega_{5}$ indicate the structural transition occur over the temperature range $60K<T<110K$ \cite{AS}.  While the FWHM of high-frequency phonon modes ($\omega_{18}$ and $\omega_{19}$) gradually broadens with an increase in temperature, the FWHM of low-frequency modes ($\omega_{1}$, $\omega_{3}$, $\omega_{4}$, and $\omega_{5}$) are very much sensitive to the temperature range $60K<T<110K$, indicating latter phonon modes are affected much by the structural phase transition. The observed temperature dependence of FWHM of all Raman modes suggests unusual electron-phonon coupling in the crystal.

\section{Conclusions}
We have investigated the anharmonicity of optical phonon modes through temperature-dependent Raman scattering measurements in $Cs_3Bi_2Br_9$. The Raman scattering measurements together with the XRD experiments reveal the contribution of volume thermal expansion to frequency shift gives the  dominant contribution to the Raman modes above 100 K. The anharmonic contributions to the frequency shift arising from one optical phonon decay into multiple acoustic phonons. The distinct anharmonic behavior below and above 100 K is interpreted as a structural phase transition at around 100 K, which is driven by the softening of the soft mode. The work demonstrates Raman spectroscopy is an excellent tool to investigate phonon anharmonicity and soft-phonon-mediated structural phase transition.

\section{Acknowledgments}

This work is supported by the Ministry of Earth Sciences, Government of India, grant number MoES/16/25/10-RDEAS. DS acknowledges the fellowship grant supported by the INSPIRE program, Department of Science and Technology, Government of India. The authors gratefully acknowledge the kind help and support from Dr. Maurizio Pplentarutti, the coordinator of XRD1 beamline, to carry out the low-temperature XRD measurements at Elettra synchrotron radiation, Trieste, Italy.

\begin{figure}[h]
	\centering
	\includegraphics[scale = 0.6]{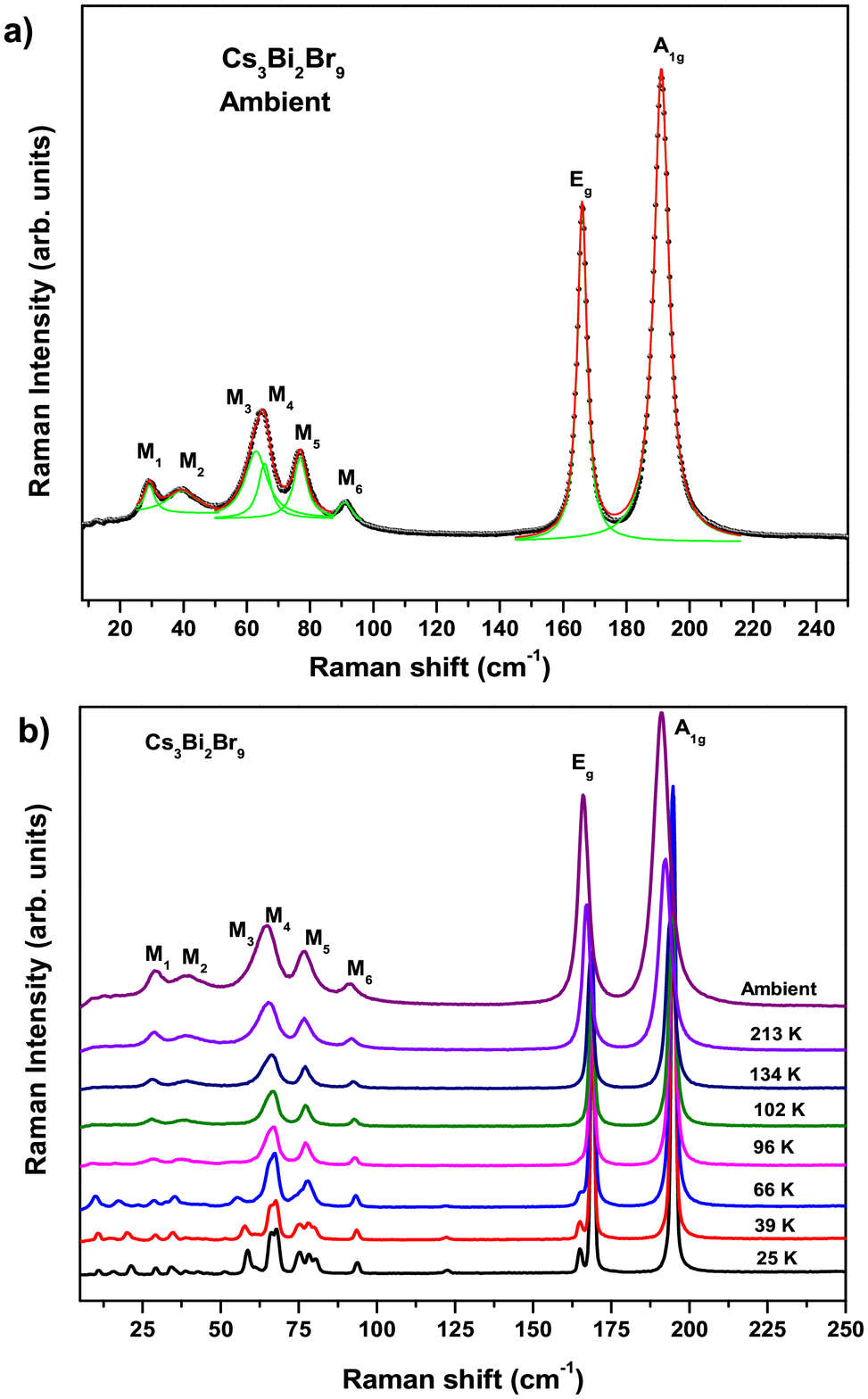}
	\caption*{FIG. 1.(a) Ambient Raman spectrum of $Cs_3Bi_2Br_9$ at room temperature.  Black dotes represent observed data points. The green line fits observed data points to Lorentzian functions and sum of that fit are shown by red line. (b) Temperature-dependent Raman spectra of $Cs_3Bi_2Br_9$ at selected temperatures.}
	
\end{figure}

\begin{figure}[h]
	\centering
	\includegraphics[scale = 0.6]{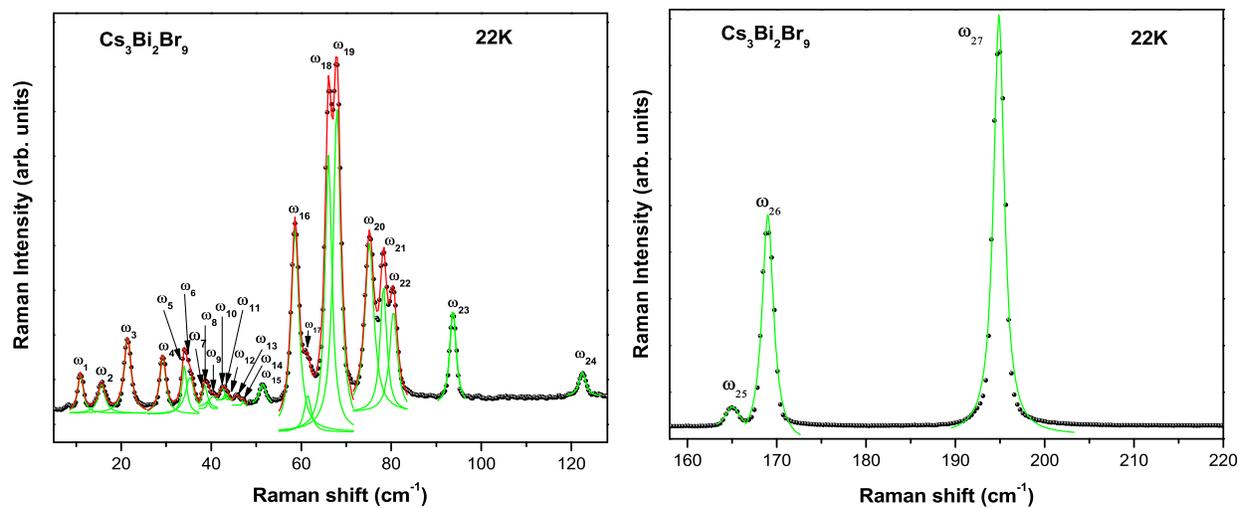}
	\vspace*{-40mm}
	\caption*{FIG. 2. Raman spectrum of $Cs_3Bi_2Br_9$ at 22 K. Black dotes represent observed data points. The green line fits observed data point to Lorentzian functions and sum of that fit are shown by red line.}

\end{figure}

\begin{figure}[h]
	\centering
	\includegraphics[scale = 0.6]{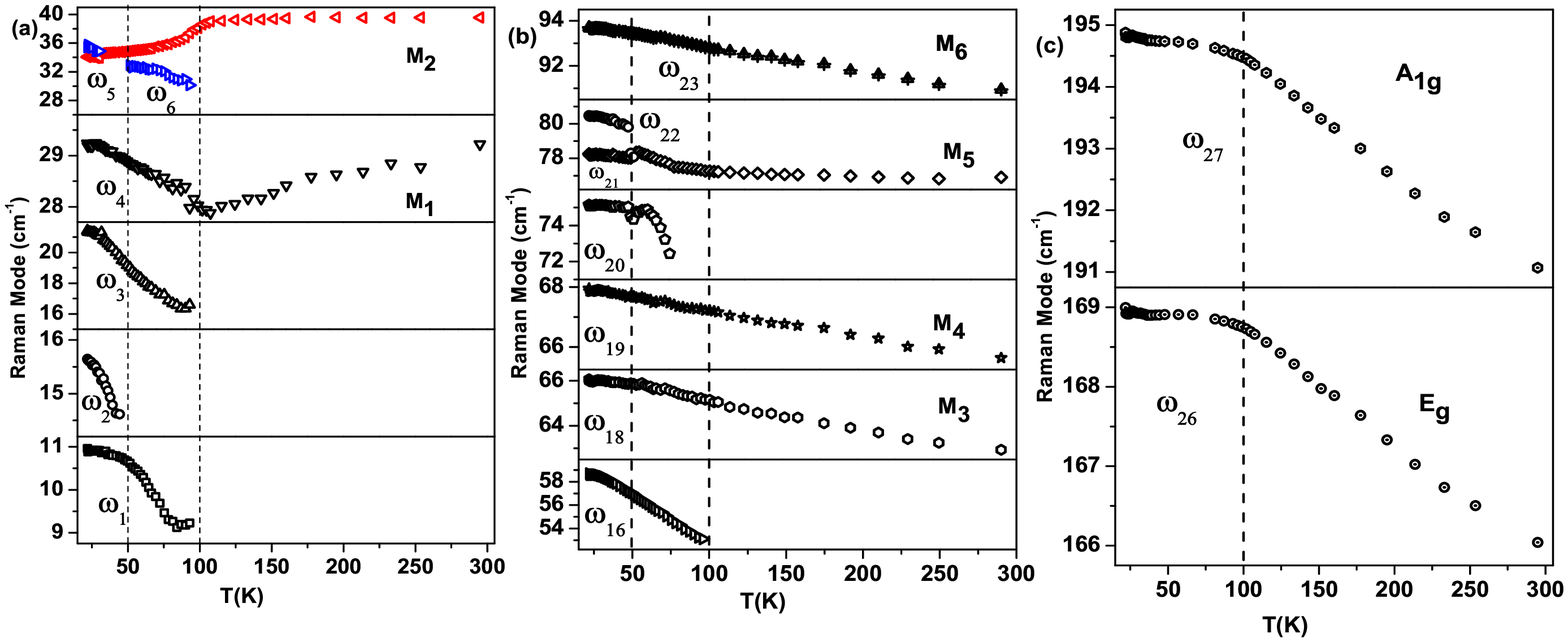}
	\vspace*{-30mm}
	\caption*{FIG. 3. Evolution of selected Raman mode frequencies with temperature.}
	\label{fig:sub1}
\end{figure}

\begin{figure}[h]
	\centering
	\includegraphics[scale = 0.5]{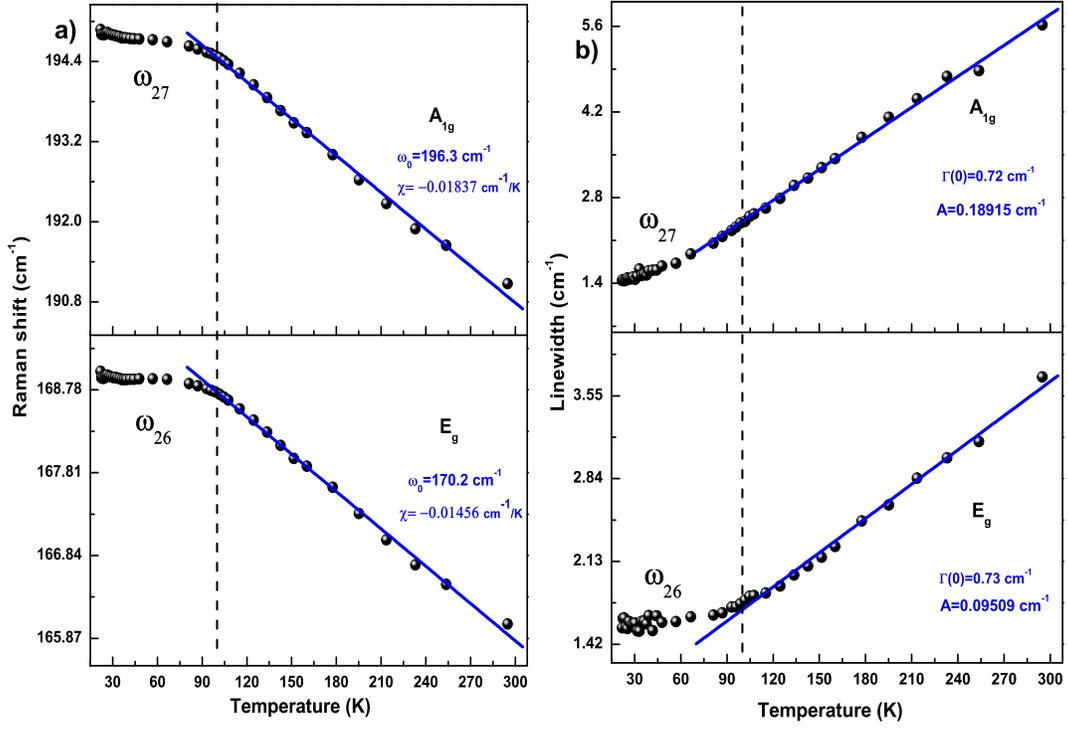}
	\caption*{FIG. 4. Variation of Raman modes (a) frequency and (b) linewidth with temperature of  $A_{1g}$ and $E_{g}$ modes. The solid line represents the fit to data using Eq. (1) as described in text.}

\end{figure}

\begin{figure}[h]
	\centering
	\includegraphics[scale = 0.5]{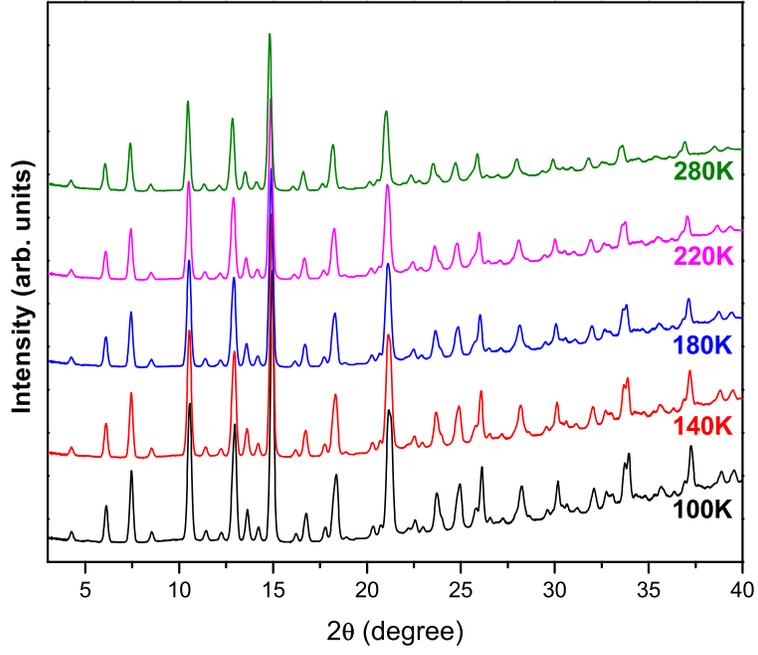}
	\caption*{FIG. 5. X-ray diffraction patterns of $Cs_3Bi_2Br_9$ at selected temperatures.}
	\label{fig:sub1}
\end{figure}

\begin{figure}[h]
	\centering
	\includegraphics[scale = 0.6]{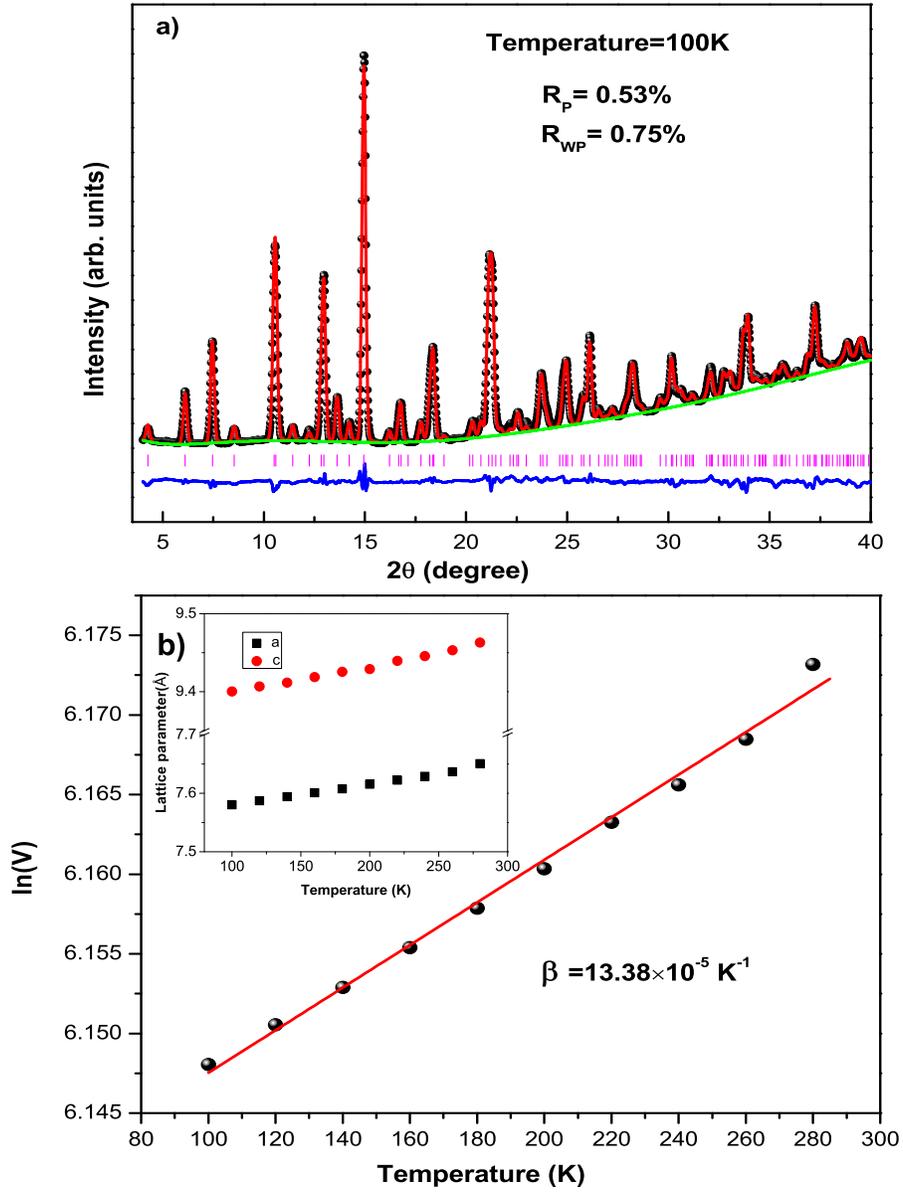}
	\caption*{FIG. 6. (a) Rietveld refinement of x-ray diffraction pattern of $Cs_3Bi_2Br_9$ obtained at 100 K. Experimental data are represented by black dote points. The red line fits to experimental data points. Background and difference are shown by green and blue line, respectively. Magenta vertical ticks indicate position of Bragg peaks of the sample. (b) A plot of ln($V$) $\it vs.$ $T$ and inset shows temperature dependence of lattice parameters.}
	
\end{figure}

\begin{figure}[h]
	\centering
    \includegraphics[scale = 0.6]{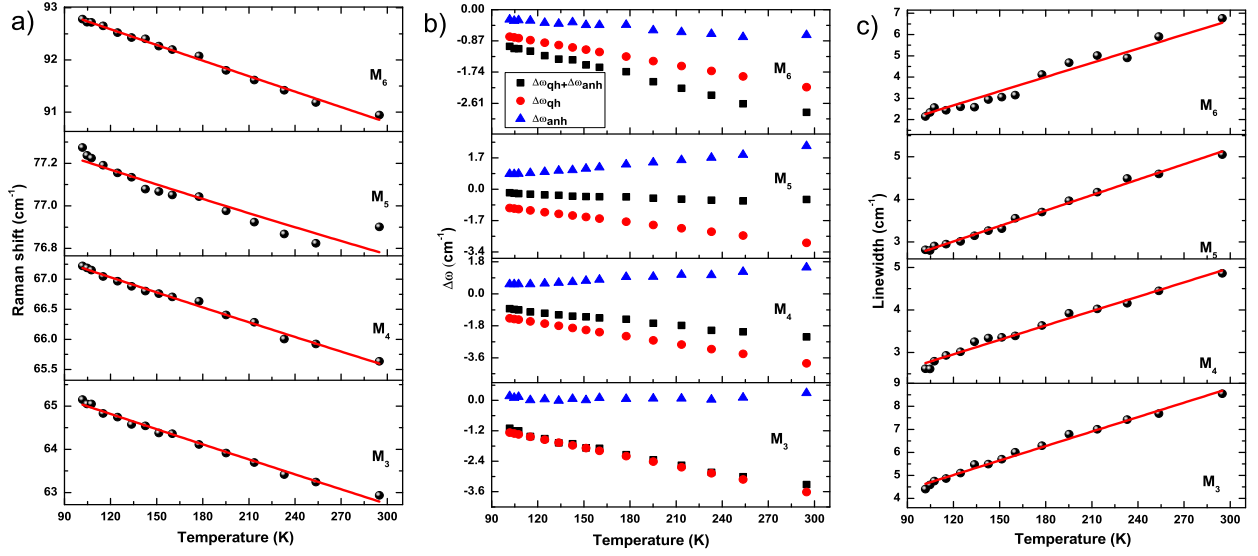}	
	\caption*{FIG. 7. (a) Variation of Raman mode frequency with temperature. (b) The frequency shift due to the volume thermal expansion $(\Delta\omega_{qh})$ and anharmonic phonon-phonon coupling $(\Delta\omega_{anh})$ and $(\Delta\omega_{qh}+\Delta\omega_{anh})$ as a function of temperature. (c) Temperature dependence of Raman linewidth for $M_{3}$, $M_{4}$, $M_{5}$, and $M_{6}$ modes. The solid line represents the fit to data using Eq. (2) and (3) as described in text.  }

\end{figure}

\begin{figure}[h]
	\centering
	\includegraphics[scale = 0.6]{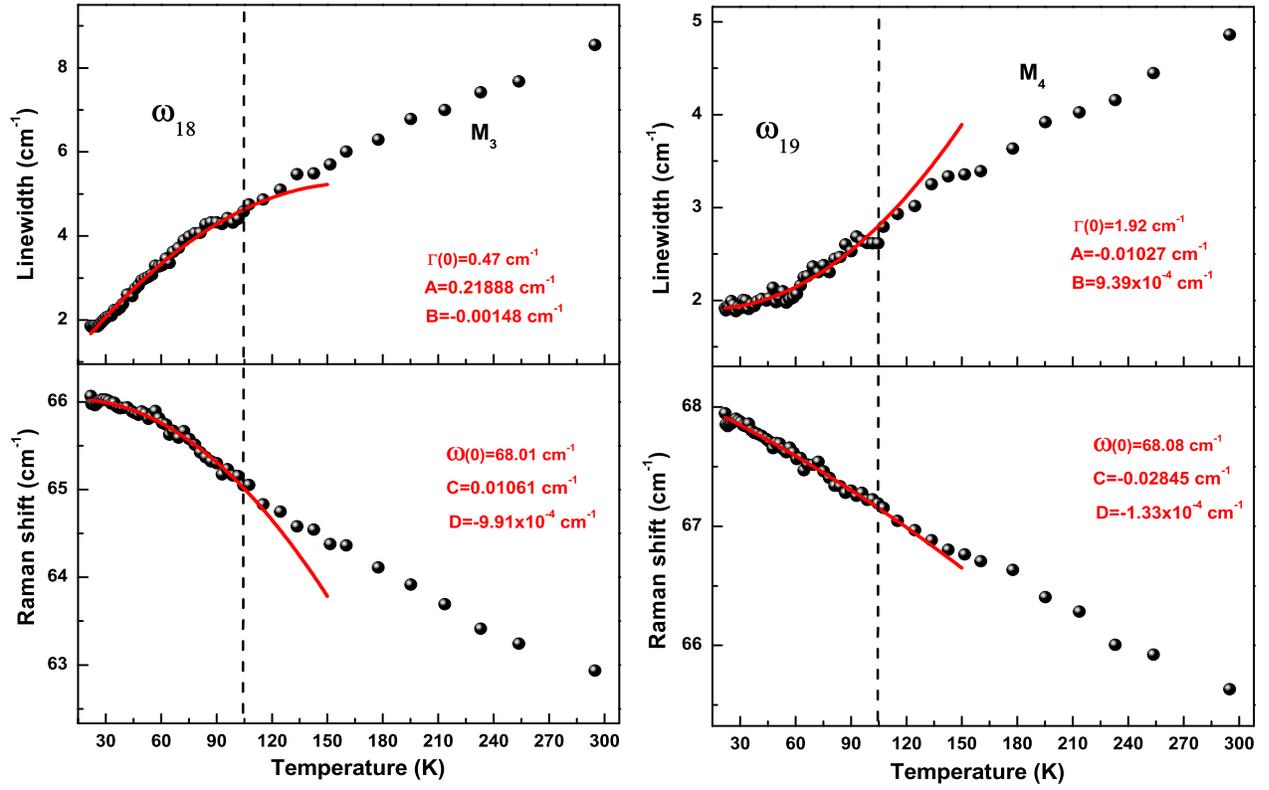}
	\caption*{FIG. 8. Evolution of Raman mode frequency and linewidth as a function of temperature for $\omega_{18}$ and $\omega_{19}$ modes. The solid line represents the fit to data using Eq. (2) and (3) as described in text.}
	
\end{figure}

\begin{figure}[h]
	\centering
	\includegraphics[scale = 0.6]{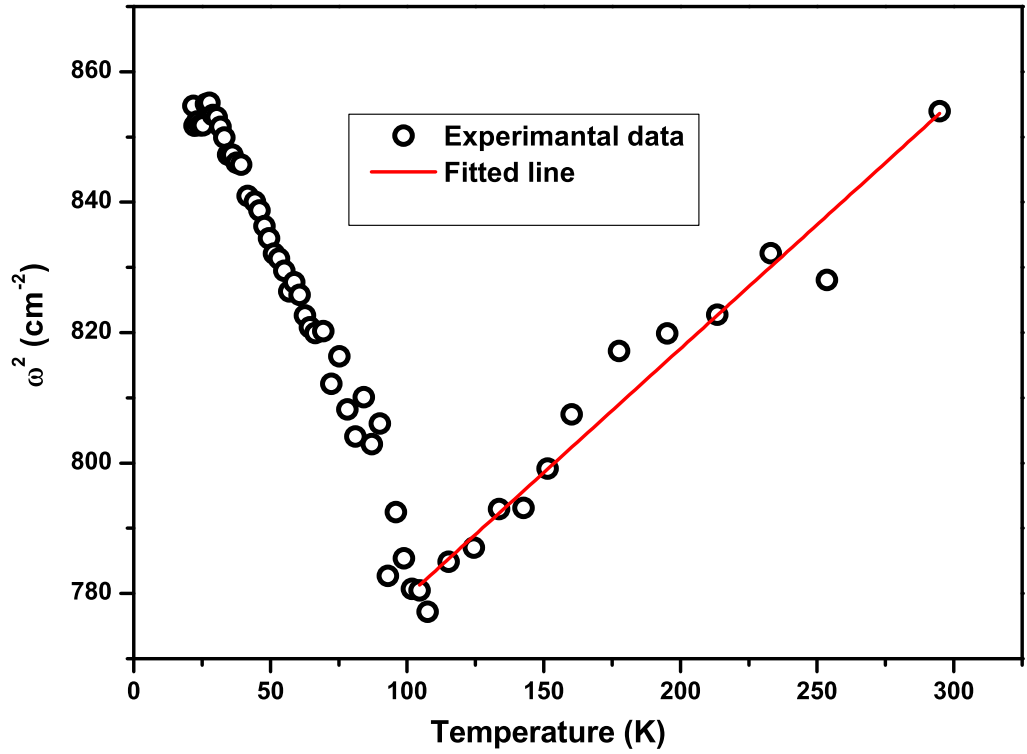}
	\caption*{FIG. 9. Evolution of a soft mode frequency with temperature.The solid line represents the fit to data using Eq. (11) as described in text.}
	\label{fig:sub1}
\end{figure}

\begin{figure}[h]
	\centering
	\includegraphics[scale = 0.6]{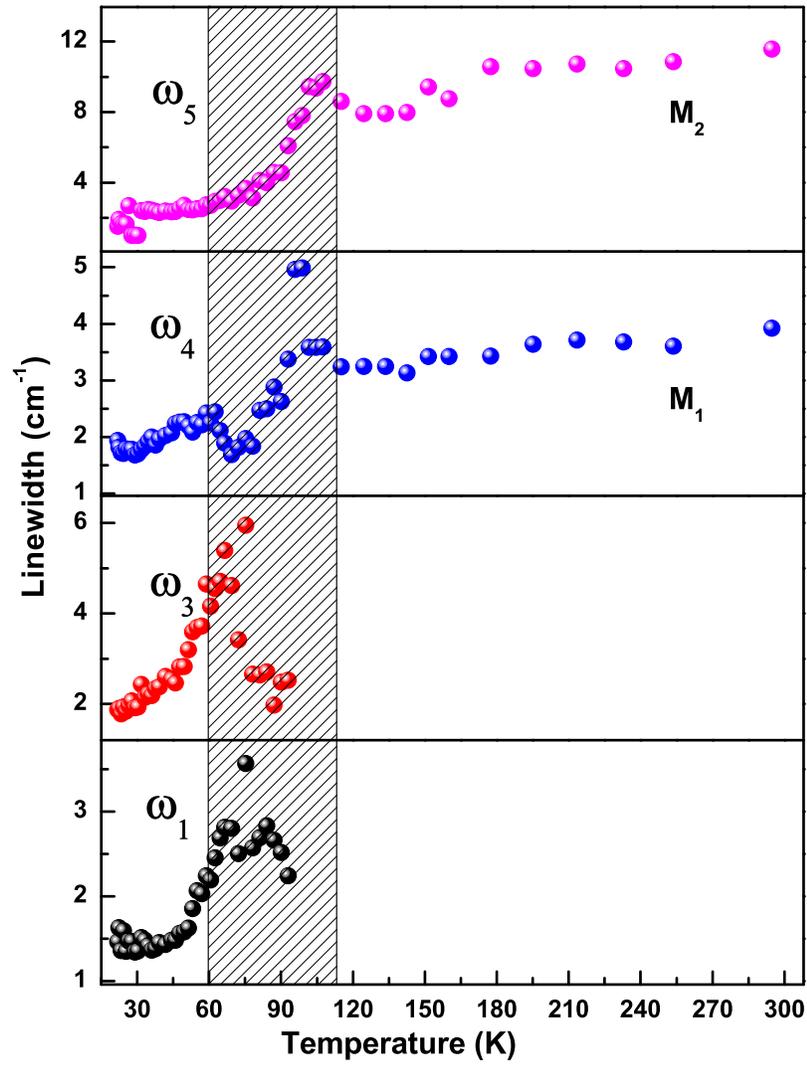}
	\caption*{FIG. 10. Evolution of linewidth of $\omega_1, \omega_3,\omega_4,  \omega_5$, $M_1$, and $M_2$ modes with temperature.}
	\label{fig:sub1}
\end{figure}

\end{document}